# Magnetization process of the $S$ = 1/2 two-leg organic spin-ladder compound BIP-BNO

Kazuya Nomura[1*], Yasuhiro H. Matsuda[1#], Yasuo Narumi[2], Koichi Kindo[1], Shojiro Takeyama[1], Yuko Hosokoshi[3], Toshio Ono[3], Naoya Hasegawa[3], Hidemaro Suwa[4,5], and Synge Todo[4,1]

[1]*Institute for Solid State Physics, University of Tokyo, Kashiwa, Chiba 277-8581, Japan*
[2]*Center for Advanced High Magnetic Field Science, Graduate School of Science, Osaka University, 1-1 Machikaneyama, Toyonaka 560-0043, Japan*
[3]*Department of Physical Science, Osaka Prefecture University, Osaka 599-8531, Japan*
[4]*Department of Physics, University of Tokyo, Tokyo 113-0033, Japan*
[5]*Department of Physics and Astronomy, University of Tennessee, Knoxville, Tennessee 37996, USA*

We have measured the magnetization of the organic compound BIP-BNO (3,5'-bis(*N-tert*-butylaminoxyl)-3',5-dibromobiphenyl) up to 76 T where the magnetization is saturated. The $S$ = 1/2 antiferromagnetic Heisenberg two-leg spin-ladder model accounts for the obtained experimental data regarding the magnetization curve, which is clarified using the quantum Monte Carlo method. The exchange constants on the rung and the side rail of the ladder are estimated to be $J_{\rm rung}/k_B$ = 65.7 K and $J_{\rm leg}/k_B$ = 14.1 K, respectively, deeply in the strong coupling region: $J_{\rm rung}/J_{\rm leg} > 1$.

1. Introduction

Spin-ladder systems are in the crossover region between one and two dimensionalities and have been attracting much attention since peculiar phenomena, such as the superconductivity in a hole doped $S$ = 1/2 two-leg spin ladder, are expected to appear.[1-4] Uehara *et al* actually observed the superconductivity in the hole-doped two-leg spin ladder $Sr_{0.4}Ca_{13.6}Cu_{24}O_{41.84}$ at 3.5 GPa.[5]

In the $S$ = 1/2 spin-ladder system with the antiferromagnetic Heisenberg (AFH) exchange interaction, it was theoretically predicted that the magnetic properties of the even-leg spin ladder and odd-leg spin ladder are essentially different; the even-leg spin ladder has a spin gap, while the odd-leg spin ladder is gapless.[6] The magnetic susceptibility of the two-leg spin-ladder compound $SrCu_2O_3$, which is the first real spin-ladder material, shows that the spin gap opens at low temperatures. On the other hand, the three-leg spin-ladder compound $Sr_2CuO_5$ exhibits the gapless feature as theoretically predicted.[7] Johnston analyzed the experimentally obtained magnetic susceptibility of $SrCu_2O_3$ in detail and estimated the spin gap and the exchange constants to be $\Delta/k_B$ = 420 K, $J_{leg}/k_B$ = 2000 K, and $J_{rung}/J_{leg}$ = 0.5, respectively,[8] where $J_{leg}$ ($J_{rung}$) is the exchange constant on the side rail (rung) of the ladder. See the model (1) below. Although the magnetic field effect on the spin gap system is intriguing and important to uncover its magnetic property, the exchange constants and resultant spin gap of $SrCu_2O_3$ seem to be too large to overcome by ordinarily obtainable magnetic fields.

These days, in addition to $SrCu_2O_3$, some other $S$ = 1/2 AFH two-leg spin-ladder compounds have been found such as $(C_5H_{12}N_2)_2CuBr_4$ (abbreviated BPCB or $(Hpip)_2CuBr_4$),[9-12] $(5IAP)_2CuBr_4 \cdot 2H_2O$[13,14], $[Cu_2(C_2O_4)(C_{10}H_8N_2)_2](NO_3)_2$,[15] $[(DT-TTF)_2][Au(mnt)_2]$,[16] and $(C_7H_{10}N)_2CuBr_4$ (abbreviated DMIPY).[17-21] $Cu_2(C_5H_{12}N_2)_2Cl_4$, or CuHpCl, is one of the most studied compounds that were considered to be spin ladders.[22-25] Some recent experiments, however, have pointed out that CuHpCl should be described as a three dimensional spin system rather than a spin ladder.[26,27] In spite of many studies on $S$ = 1/2 spin-ladder systems both theoretically and experimentally, there are few compounds with exchange constants ($J_{rung}$ and $J_{leg}$) comparable to readily available magnetic field ($B$): $B$ < 20 T. BPCB, $(5IAP)_2CuBr_4 \cdot 2H_2O$, and DMIPY are the materials that have such weak exchange constants and the ratios of the exchange constants $J_{rung}/J_{leg}$ were estimated to be approximately 3.5, 13, and 0.52, respectively.[10, 13, 17] The magnetic field effects on these materials have been extensively investigated. It is found that the Tomonaga-Luttinger liquid (TLL) phase appears when the spin gap closes at high magnetic fields.[11, 12, 19, 21]

In the meantime, the existence of an $S$ = 1/2 two-leg spin-ladder material without transition metal element has not been confirmed. Although the susceptibility measurement of BIP-BNO, where NO groups have $S$ = 1/2 spins,[28] is reasonably consistent with such a spin-ladder model, it is hard to distinguish a spin ladder from a bond-alternating chain by means of the susceptibility measurement. Meanwhile, it is expected that another organic compound, $[(DT-TTF)_2][Au(mnt)_2]$, is also described by a spin-ladder model,[16] but there has been no experimental evidence of its realization, because the strong exchange interaction makes it difficult to study its magnetization process.

The key to the identification lies in the magnetization process. In the previous study of BIP-BNO, the magnetization process was reported up to 50 T.[29] The magnetization curve shows the non-magnetic ground state below the critical point $B_{C1}$ = 35 T and begins to grow at $B_{C1}$. By fitting the experimental data to a ladder model using exact diagonalization, they estimated $J_{\text{leg}}/k_B = 28$ K and $J_{\text{rung}}/k_B = 73.5$ K. However, at 50 T the observed magnetization is approximately only a quarter of the saturation value $M_{\text{sat}} = 2$ $\mu_B$/f.u. To clarify the property of BIP-BNO as an $S = 1/2$ two-leg spin-ladder system and to determine accurate exchange constants, it is required to analyze the full magnetization curve and see how well the ladder model can explain the experiment. Our purpose of the present study is to provide an unambiguous evidence that BIP-BNO is well described by a spin-ladder model by the measurement of magnetization in high fields up to 76 T.

## 2. Experiment

To generate high field magnetic fields, we used the non-destructive type multilayer pulsed magnet (MLPM)[30] and the destructive type vertical single-turn coil (VSTC)[31] at the International Mega Gauss Science Laboratory of the Institute for Solid State Physics, the University of Tokyo. The duration time of the magnetic field generated by the MLPM used in this study is about 10 ms and that by the VSTC is about 7 ms. For magnetization measurements, induction method using coaxial type or parallel-type pick up coil was employed. The signal from pick up coil is proportional to time ($t$) derivative of the magnetization ($M$), $dM/dt$. Integrating the measured signal as a function of $t$, the magnetization curve is obtained.

The single crystals of BIP-BNO were grown by recrystallization from a concentrated solution of $CH_2Cl_2$ in acetonitrile atmosphere at -10°C. We used microcrystalline sample and single crystals for the non-destructive MLPM experiment and the destructive VSTC experiment, respectively. In VSTC measurements, about ten single crystals were aligned such that the c-axis of each crystal was parallel to the applied magnetic fields.

## 3. Numerical simulation

To compare experimental data with the theoretical model, we calculated the magnetization of the spin-ladder system under a magnetic field by means of the worldline quantum Monte Carlo method with the worm (directed-loop) algorithm. The worm scattering probability is optimized using the geometric allocation[32, 33]. The exchange constants of the model (1) below are estimated using the simulated annealing in which the deviation from the experimentally obtained magnetization curve is minimized. We confirmed that the results of 64 and 128 spins are consistent with each other within the error bars. Thus, in the present paper, we show the magnetization of 128 spins as the value in the thermodynamic limit.

## 4. Results and Discussion

In Fig.1, the triangles represent the magnetizations of the microcrystalline sample measured with MLPM, and the circles represent those of the single crystal measured with VSTC. Although the pulse duration of VSTC is three orders of magnitude shorter than that of MLPM, the obtained magnetization curves with these two different methods are found to be almost identical in the field range up to 70 T, which suggests an accurate measurement was also made with the destructive method using VSTC. The error of the magnetization with MLPM is smaller than the triangle mark, and the error with VSTC is approximately 3 %, which is almost the same size of the circle mark in Fig. 1.

Below 37 T, the ground state has zero magnetization, which is consistent with the previous study[29]. The magnetization starts increasing at 37 T and shows the saturation at fields exceeding 74 T.

Since the microcrystalline sample and the single crystal show almost the same magnetization curves, BIP-BNO as well as other compounds composed of NO groups[34, 35] is expected to have no spin anisotropy.

We consider the simple $S = 1/2$ AFH two-leg spin-ladder model under magnetic field is considered:

$$\mathcal{H} = J_{\text{leg}} \sum_i (S_{1,i} S_{1,i+1} + S_{2,i} S_{2,i+1}) + J_{\text{rung}} \sum_i S_{1,i} S_{2,i} - \mu_B g B \sum_i (S_{1,i} + S_{2,i}), \quad (1)$$

where $i$ denotes the site index, $g$ is the $g$-factor assumed to be 2 in this study, and $m_B$ is the Bohr magneton.

Setting temperature to be 2 K, we estimate $J_{\text{rung}}/k_B = 65.7$ K and $J_{\text{leg}}/k_B = 14.1$ K by means of the quantum Monte Carlo method and the simulated annealing; that is, the ratio $J_{\text{rung}}/J_{\text{leg}}$ is estimated to be approximately 4.7. The spin-ladder model with the estimated exchange

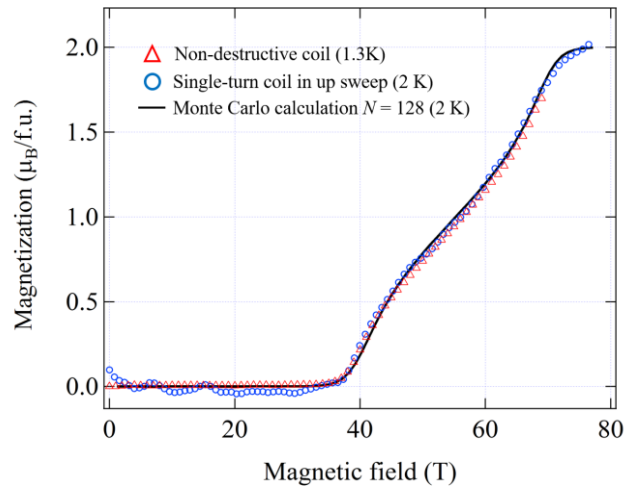

Fig.1 Magnetization curve of BIP-BNO. The circles and triangles represent the experimental results of the single turn coil and the non-destructive coil, respectively. The solid line shows the numerical result calculated by the quantum Monte Carlo method.

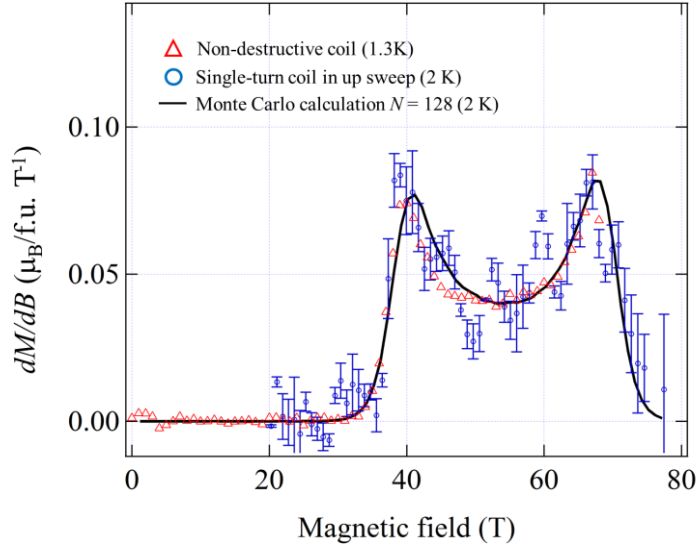

Fig. 2 Magnetic field derivative of magnetization.

constants reproduces well the whole magnetization curve obtained in the experiments as shown in Fig. 1. Figure 2 represents the magnetic field derivative of the magnetization ($dM/dB$). The solid line, the circles, and the triangles represent the numerical calculation, the experimental result with VSTC, and that with MLPM, respectively. The error bars represent the statistical errors of the experiments by VSTC. The sharp two peaks are clearly seen in the numerically calculated as well as experimentally obtained $dM/dB$ curves. We expect that the phase in the region where the magnetization is larger than zero and smaller than the saturation value should be the TLL as it is suggested in other two-leg spin-ladder materials.[11, 12, 19]

Moreover, the inflection point between the two peaks is located at approximately 55 T where $M = M_{sat}/2$, and the symmetric form with respect to the inflection point is clearly seen. Although it is generally known that we cannot distinguish between a spin ladder and a bond-alternating chain from the temperature dependence of the magnetic susceptibility,[9, 10] the magnetic field dependence of $dM/dB$ is distinguishable. Two peaks in the $dM/dB$ curve of a bond-alternating chain have different magnitudes. The magnitude of the first peak is from 20 % to 70 % as large as that of the second peak, depending on the alternation parameter. And the inflection point between two peaks is not at the center position.[36] On the other hand, a symmetric shape in $dM/dB$ is, in general, expected for a two-leg spin ladder.[25] As we have seen, BIP-BNO shows only 3% difference between the magnitudes of the two peaks in $dM/dB$ and the inflection point located at the center between the two peaks, which cannot be attributed to a bond-alternating chain. These findings provide strong evidence that BIP-BNO is identified with the $S = 1/2$ two-leg spin ladder.

## 5. Conclusion

In this study, we have conducted the magnetization measurements up to 76 T on the organic $S$

= 1/2 magnet BIP-BNO. The quantum Monte Carlo calculation of the AFH two-leg spin ladder reproduces the experimentally obtained magnetization curve of BIP-BNO. The ratio of the Heisenberg exchange constant $J_{rung}/J_{leg}$ is estimated to be approximately 4.7, deeply in the strong coupling region: $J_{rung}/J_{leg} > 1$. In the magnetic field derivation of the magnetization, *dM/dB*, the characteristic features, namely two sharp peaks and a centered inflection point, are observed in the symmetric shape of the structure. This observation strongly suggests that BIP-BNO is identified with the *S* = 1/2 AFH two-leg spin ladder. It is also worth noting that BIP-BNO is the first prototypical organic (not containing magnetic ions) compound of the spin ladder.

## Acknowledgment

This work was partly supported by JSPS KAKENHI, Grant-in-Aid for Scientific Research (B) (Grant No. 16H04009).

*knomura@issp.u-tokyo.ac.jp  
#ymatsuda@issp.u-tokyo.ac.jp